# Analogous Patterns of Student Reasoning Difficulties in Introductory Physics and Upper-Level Quantum Mechanics

Chandralekha Singh and Emily Marshman

*Department of Physics and Astronomy, University of Pittsburgh, Pittsburgh, PA, 15260, USA*

**Abstract**: Very little is known about how the nature of expertise in introductory and advanced courses compares in knowledge-rich domains such as physics. We develop a framework to compare the similarities and differences between learning and patterns of student difficulties in introductory physics and quantum mechanics. Based upon our framework, we argue that the qualitative patterns of student reasoning difficulties in introductory physics bear a striking resemblance to those found for upper-level quantum mechanics. The framework can guide the design of teaching and learning tools.

## INTRODUCTION AND FRAMEWORK

It is widely assumed that a majority of upper-division physics students have learned not only a larger amount of physics content, but have also developed significantly better reasoning, problem solving and meta-cognitive skills than introductory physics students. However, expertise is domain-specific—it is unclear how readily skills transfer across domains [1-3]. Classical mechanics (CM) and quantum mechanics (QM) are two significantly different paradigms. Learning QM can be challenging even for advanced students who have developed a good knowledge structure of CM [4,5]. Here, we develop a framework and argue that the level of challenge that advanced students face in learning the new paradigm of QM is comparable to what many students face in learning introductory physics.

Physics is a knowledge-rich domain, and the laws of physics are encapsulated in precise mathematical forms. While the level of pre-requisite mathematical knowledge is different (e.g., trigonometry, algebra or basic calculus for various introductory physics courses vs. linear algebra and advanced calculus including differential equations for QM), in both introductory physics and QM, students must learn to unpack the compact mathematical laws of physics and apply them in diverse situations to explain and predict physical phenomena. Taking into account the required prior knowledge of students learning introductory physics or QM, each group must learn to interpret and make sense of abstract physical principles and make a conscious effort to build a coherent knowledge structure in order to become an expert [1-3]. Once we correct for the pre-requisite prior knowledge of students in each case, introductory physics content for a beginning student is no less abstract than QM content is for an advanced physics student.

Since students are not blank slates, the difficulty of introductory physics courses is increased. Students constantly try to make sense of the world around them. The mental models they build of how things work in everyday life using naïve reasoning based upon their limited expertise are often inconsistent with the laws of physics [6]. Moreover, everyday terms such as velocity, acceleration, momentum, energy, work etc. do not have the same precise meaning as in physics and students must learn to differentiate how those terms are used in physics vs. everyday life.

Students are unlikely to have unproductive mental models about QM concepts before formal instruction in physics because one does not encounter such situations and need to reason about quantum processes in everyday life. Therefore, one might assume that learning QM may be easier than CM in this regard. However, what students learn in earlier courses including CM can interfere with building a robust knowledge structure of QM. For example, in QM, the connection between quantum formalism and phenomena is made through measurement and inferences about physical observables, e.g., position, momentum, energy, angular momentum. But unlike CM, a particle does not in general have a definite position, momentum or energy in QM. In QM, all information about the system is contained in the state vector or wave function which lies in an abstract vector space. The measurement of an observable collapses the wave function to an eigenstate of the operator corresponding to the observable measured, and the probability of measuring a particular value can be calculated from the knowledge of the wave function. In fact, this QM paradigm is so different from the classical paradigm that students must build a knowledge structure for QM from scratch, even if they have a good knowledge structure involving position, momentum, and angular momentum in CM. Moreover, similar to the possibility of naïve notions about velocity, momentum or work from everyday experience interfering with learning CM, concepts of position, momentum, angular momentum, etc. are embedded so differently in CM and QM formalisms that intuition about these concepts developed in CM may actually interfere with learning QM.

Within our framework, the significantly different CM and QM paradigms suggest that even students with a good knowledge of CM will start as a novice and gradually build their knowledge structure about QM. As these students start to build a knowledge structure about QM, their knowledge will initially be in disconnected pieces [6], and their reasoning about QM will only be locally consistent and lack global consistency. In fact, there is nothing unusual about students going through this stage. Those who begin their pursuit of developing expertise in any knowledge-rich domain must go through a phase in which their knowledge is in small disconnected pieces which are only locally consistent, and this causes reasoning difficulties [1-3]. What our framework suggests, however, is that each student must go through this process of gradually building a knowledge structure and pass through the knowledge in pieces phase while learning CM and QM separately because the conceptual frameworks are sufficiently different in these sub-domains of physics (even though the same terminology is used, e.g., momentum, energy, etc). Thus, within our framework, students' mastery of CM does not imply their mastery of QM without conscious effort on the part of the students to build a knowledge structure of QM (and make lateral connections between the CM and QM schema to explicitly understand the differences between these formalisms and when and how they come together, e.g., by taking the classical limit). Therefore, students learning CM and QM are likely to show similar patterns of reasoning difficulties as they move up along the expertise spectrum in each of these sub-domains of physics. In each case, if students continue their effort to repair, reorganize and extend their knowledge structure [1-3], they will reach a point when their knowledge structure becomes robust, and they are able to make predictions and inferences which are globally consistent with the respective formalisms and their reasoning difficulties are significantly reduced.

## PATTERNS OF DIFFICULTIES

Below, we discuss a few examples from upper-level QM and introductory CM to illustrate that the patterns of student difficulties are analogous in these cases. As discussed in the framework above, in each case, students are still learning to unpack the respective principles and formalism encapsulated in mathematical forms and developing schema.

We first discuss the performance of 39 physics graduate students on questions related to QM formalism given as part of a written test after at least one semester of graduate-level QM.

On question 1 which asked graduate students whether the statement *"The position-space wavefunction is $\psi(x) = \langle x|\psi\rangle$ where x is a continuous index"* is correct, 90% noted that it is correct. However, responses to other questions in the same test, e.g., question 2 below, suggest that many students do not recognize that $\psi(x) = \langle x|\psi\rangle$ in other contexts.

2. *An operator $\hat{Q}$ corresponding to a physical observable in the position representation is $Q(x)$. Choose all of the following statements that are correct.*
1. $\langle x|\hat{Q}|\psi\rangle = Q(x)\langle x|\psi\rangle$  2. $\langle x|\hat{Q}|\psi\rangle = Q(x)\psi(x)$
3. $\langle x|\hat{Q}|\psi\rangle = \langle \psi|\hat{Q}|x\rangle$
*A. 1 only   B. 2 only   C. 1 and 2 only   D. 1 and 3 only   E. 2 and 3 only*

**TABLE 1**. Percentages of graduate students who selected various answer choices for questions related to QM (correct answers are italicized). Percentages do not add up to 100% since some students left it blank.

|            | A  | B  | C      | D  | E  |
|------------|----|----|--------|----|----|
| Question 2 | 13 | 10 | *54*   | 8  | 10 |
| Question 3 | 26 | 13 | *46*   | 13 | 0  |

Table 1 shows that for question 2, 41% of graduate students incorrectly claimed that either statement (1) or (2) is correct but not both, even though in the earlier question, 90% of them selected $\psi(x) = \langle x|\psi\rangle$ as correct as noted earlier (interpreting $\psi(x) = \langle x|\psi\rangle$ is the only difference between statements (1) and (2) in question 2). In fact, there were two more questions in the same test (which were placed back to back with the two questions discussed above) which show similar discrepancies. In one of those questions, 36% of the graduate students incorrectly claimed that $\langle x|\psi\rangle = \int x\psi(x)dx$ is correct and 39% incorrectly claimed that $\langle x|\psi\rangle = \int \delta(x-x')\psi(x')dx'$ is not correct.

These discrepancies indicate that graduate students are inconsistent in their reasoning and their ability to correctly identify $\psi(x) = \langle x|\psi\rangle$ is context-dependent. Interviews in which graduate students reasoned through these problems using a think aloud protocol suggest that they did not reflect back and check for consistency in their responses to the four consecutive questions. Assuming a graduate student has self-monitoring skills, inconsistency across consecutive questions suggests that either the student did not feel the need to check for consistency or he experienced cognitive overload due to limited capacity of working memory. Thus, the student may be unable to do self-monitoring while solving these problems in a domain in which he is still developing expertise [1-3]. Interviews suggest that in question 2, students were so focused on the operator and how it should be written in position representation, that they did not pay attention to the fact that the only difference between statements 1 and 2 was $\psi(x) = \langle x|\psi\rangle$, which they had already identified as correct previously. A similar inconsistency was observed when Siegler [8] taught young children control of variables for a balance scale in a longitudinal study. He found that after a few rounds of instruction, children were in an intermediate state of expertise and used the control of variables correctly on some days on some tasks but not in others.

In the interview, a graduate student who noted that $\psi(x) = \langle x|\psi\rangle$ in question 1 but incorrectly claimed that $\langle x|\psi\rangle = \int \delta(x-x')\psi(x')dx'$ is incorrect reasoned as follows "… it just doesn't seem correct, that $\psi(x)$ should just pop out [of the integral]. It's giving you just a wavefunction of $x$ and I just don't like that. I think it [inner product] should just give you a number." He correctly reasoned that the inner product is a number, but did not make the connection that $\psi(x)$ is also a number for any particular value of $x$. More importantly, he was so focused on his concern that the inner product is a number that he did not notice the inconsistency between the responses to this question and question 1 in which he appeared quite confident that $\psi(x) = \langle x|\psi\rangle$. We note that an integral involving a delta function of the type shown above is trivial for a graduate student if it is given as a math problem. However, in the context of QM, that integral involving a delta function was enough to make this student (and many others) concerned about whether the physical content of that statement made sense from the point of view of QM when it was nothing more than $\psi(x) = \langle x|\psi\rangle$, whose validity he appeared confident about in the previous question.

This type of lack of consistency is well-documented in introductory physics. For example, a student may correctly reason in a simple context that a larger net force on an object would imply a larger acceleration, but incorrectly claim that the net force is larger on a car moving at a constant velocity at 100 mph compared to one that is moving at 60 mph.

Another type of difficulty is illustrated by the following QM question given to the same 39 graduate students.

*3. Consider the following conversation between Andy and Caroline about the measurement of an observable Q for a system in a state $|\psi\rangle$ which is not an eigenstate of $\hat{Q}$:*

*Andy: When an operator $\hat{Q}$ corresponding to a physical observable Q acts on the state $|\psi\rangle$, it corresponds to a measurement of that observable. Therefore, $\hat{Q}|\psi\rangle = q_n|\psi\rangle$ where $q_n$ is the observed value.*

*Caroline: No. The measurement collapses the state so $\hat{Q}|\psi\rangle = q_n|\psi_n\rangle$ where $|\psi_n\rangle$ on the right hand side of the equation is an eigenstate of $\hat{Q}$ with eigenvalue $q_n$.*

*With whom do you agree?*
*A. Agree with Caroline only, B. Agree with Andy only,*
***C. Agree with neither**, D. Agree with both,*
*E. The answer depends on the observable.*

Table 1 shows that for question 3, 52% of the graduate students incorrectly agreed either with Andy or Caroline or both. Interviews suggest that students were so focused on thinking about how a single equation should describe the measurement process and the collapse of the wave function under the Copenhagen interpretation, that none of them felt the need to check and verify that equations such as $\hat{Q}|\psi\rangle = q_n|\psi_n\rangle$ are meaningless from the point of view of linear algebra. Graduate students are unlikely to make such mistakes if a "pure" linear algebra question is asked regarding the validity of similar equations without the QM context. However, in this context involving quantum measurement, their misconception that an operator (corresponding to an observable) acting on a quantum state corresponds to the measurement of the observable was so strong that they did not pay attention to the violation of basic tenets of linear algebra. In fact, some of the interviewed students who claimed that both Andy and Caroline are correct (which again makes no sense from a mathematical point of view since the left hand sides are the same but the right hand sides are different in Andy's and Caroline's equations), when pressed for an explanation for how only the right hand side of an equation would change, noted that Andy is correct infinitesimally before the measurement process occurs and Caroline is correct infinitesimally after it.

Similar overlooking of mathematical or other types of consistency due to strong misconceptions is common in introductory physics. For example, we gave to introductory students two isomorphic problems [9] involving Newton's second law in an equilibrium situation on an incline plane back-to-back (problems that can be mapped onto each other but the contexts were different). The second problem elicits a strong misconception about static friction always taking on its maximum value. We found that many introductory students ignored the similarity between the adjacent problems (including the fact that the free-body diagrams provided were identical except that the tension force in one problem was replaced by the frictional force in the other problem, which would logically imply that the desired quantities, tension and friction, had the same magnitudes). Only 28% of the students provided the correct response to the friction problem (72% correct for tension problem right before the friction problem). A majority did not recognize a similarity between the isomorphic problems—despite doing the tension problem correctly—and launched into a calculation of maximum static friction although it was not at the maximum value in the problem.

Another common difficulty with QM is manifested by the fact that 48% of the graduate students incorrectly claimed that $\hat{H}|\psi\rangle = E|\psi\rangle$ is the most fundamental equation of QM and 39% claimed it is true for all wavefunctions [5]. In individual interviews, students were explicitly asked whether this equation is true for a linear superposition of the ground and first excited states of a one dimensional infinite square well. Many graduate students incorrectly claim that it is indeed true for that case primarily because they believe incorrectly that this time-independent Schrodinger equation is the most fundamental equation of QM. When these students are asked to explicitly show that this equation is true in this given context, most of them verbally argue without writing that $\hat{H}|\psi_1\rangle = E_1|\psi_1\rangle$ and $\hat{H}|\psi_2\rangle = E_2|\psi_2\rangle$ imply that their addition will give $\hat{H}|\psi\rangle = E|\psi\rangle$. In fact, even when graduate students are told that $\hat{H}|\psi\rangle = E|\psi\rangle$ is not obtained, many have difficulty believing until they explicitly write these equations on paper (mostly after additional encouragement to do so) and note that since $E_1$ and $E_2$ are not equal, $\hat{H}|\psi\rangle \neq E|\psi\rangle$ in this case.

Introductory physics students show similar patterns of difficulty in that they use their "gut" feeling to answer physics questions instead of explicitly writing down a physics principle and checking its applicability in a particular situation. For example, we performed a study [9] in which 138 introductory students were asked to find a mathematical expression (*Ft*) for the magnitude of the momentum of a boat that started from rest and had a constant horizontal force of magnitude *F* acting on it for a time *t* (and in which the force was used to tow the boat a distance *d*). Another group of 215 introductory students were asked a similar but conceptual question in which two boats started from rest and had the same constant net horizontal force acting on each for the same period of time. They were asked to compare the magnitudes of momentum of each boat. Both the quantitative and conceptual questions were in the multiple-choice format. Many introductory students used their gut feeling rather than physics principles to answer this question. The percentage of students providing the correct response for the conceptual question was roughly half of the percentage of students who correctly answered the quantitative problem. When a third group of 289 students (different from the first two) was given both questions with the quantitative question followed by the conceptual question, they performed equally well on both. Interviews suggest that introductory students who solved the quantitative problem took advantage of their expression (*Ft*) to answer the conceptual question. However, during interviews, introductory students who were only given a conceptual question were very reluctant to convert it into a quantitative problem in order to answer it [7] (similar to the reluctance of the graduate students who preferred to use their gut feeling to argue that $\hat{H}|\psi\rangle = E|\psi\rangle$ for the case involving a linear superposition of stationary states, even though it is not true).

In our framework, the reluctance of introductory students to use their cognitive resources for quantitative analysis of conceptual questions is similar to the reluctance of graduate students to verify the validity of $\hat{H}|\psi\rangle = E|\psi\rangle$ explicitly by writing it down in the given situation until they were forced to do it. One possible explanation for such reluctance is that writing down each step explicitly and converting a conceptual question to a quantitative question in order to solve it are cognitively demanding tasks and may cause mental overload [7]. According to Simon's theory of bounded rationality [2], an individual's rationality in a particular context is constrained by his expertise and experience and an individual will choose amongst only a few options consistent with his expertise that does not cause a cognitive overload.

## SUMMARY

We develop a framework and argue that the patterns of student difficulties in introductory physics are analogous to the patterns in advanced QM. The novel paradigm in QM requires that even those who have mastered CM construct a new knowledge structure and gradually make their way up the expertise spectrum similar to the process that an introductory student must go through to learn introductory physics.

## ACKNOWLEDGEMENT

We thank the National Science Foundation.